\documentclass[fleqn,usenatbib]{mnras}

\usepackage{graphicx}	% Including figure files
\usepackage{amsmath}	% Advanced maths commands
\usepackage{amssymb}	% Extra maths symbols

\usepackage{color}
\usepackage[dvipsnames]{xcolor}

\def\fsc{\alpha_\mathrm{f}}                              
\def\omegaB{\omega_{\mathrm{B}}}
\def\sigt{\sigma_{\mathrm{T}}}

\title[Radiative transfer simulations of magnetar flare beaming]{Radiative transfer simulations of magnetar flare beaming}

\author[T. van Putten et al.]{
T. van Putten,$^{1}$\thanks{E-mail: T.vanPutten@uva.nl}
A. L. Watts,$^{1}$
M. G. Baring,$^{2}$
and R. A. M. J. Wijers$^{1}$
\\
$^{1}$Anton Pannekoek Institute for Astronomy, University of Amsterdam, Postbus 94249, 1090 GE Amsterdam, The Netherlands\\
$^{2}$Department of Physics and Astronomy, Rice University, 6100 Main St., Houston, TX, 77005-1892, USA\\
}

\date{Accepted XXX. Received YYY; in original form ZZZ}

\pubyear{2016}

\begin{document}
\label{firstpage}
\pagerange{\pageref{firstpage}--\pageref{lastpage}}
\maketitle

\begin{abstract}
Magnetar giant flares show oscillatory modulations in the tails of their light curves, which can only be explained via some form of beaming. The fireball model for magnetar bursts has been used successfully to  fit the phase-averaged light curves of the tails of giant flares, but so far no attempts have been made to fit the pulsations. We present a relatively simple numerical model to simulate beaming of magnetar flare emission.  In our simulations, radiation escapes from the base of a fireball trapped in a dipolar magnetic field, and is scattered through the optically thick magnetosphere of the magnetar until it escapes. Beaming is provided by the presence of a relativistic outflow, as well as by the geometry of the system. We find that a simple picture for the relativistic outflow is enough to create the pulse fraction and sharp peaks observed in pulse profiles of magnetar flares, while without a relativistic outflow the beaming is insufficient to explain giant flare rotational modulations. 
\end{abstract}

\begin{keywords}
radiative transfer -- stars: magnetars -- X-rays: bursts
\end{keywords}

%%%%%%%%%%%%%%%%%%%%%%%%%%%%%%%%%%%%%%%%%%%%%%%%%%

%%%%%%%%%%%%%%%%% BODY OF PAPER %%%%%%%%%%%%%%%%%%

\section{Introduction}

Magnetars are neutron stars with a very strong inferred magnetic field \citep{Duncan1992,Paczynski1992a}, that commonly emit bursts with photon energies ranging between a few and a few hundred keV. These bursts are usually divided into three classes by their energetics: short bursts with total energy up to $10^{41}$ erg, intermediate flares with energy $10^{41-43}$ erg and the very rare giant flares with total burst energies of $10^{44-46}$ erg \citep[see][for reviews]{Woods2006a, Turolla2015a}.

The light curves of magnetar bursts vary wildly in shape and duration. Most have a sharp initial rise with a strong peak of emission, which can then be followed by a slowly decaying tail. Most short bursts do not have an observable tail, but many of the intermediate flares do have this tail, as do all three known giant flares. The tails of the intermediate and giant flare light curves show modulations at the rotation frequency of the magnetar. These modulations tend to be strong in the giant flares, up to an order of magnitude in amplitude \citep[see for example][]{Hurley2005a}, but are weaker for intermediate flares. They can consist of multiple pulses per rotation period, as in the case of the 1998 giant flare \citep[][four subpulses]{Hurley1999a} and the 2004 giant flare \citep[][two subpulses]{Hurley2005a}. This suggests the emission making up the tail of a magnetar burst has a preferred direction, either through the physical location of the emitting region or via some form of beaming.

The trigger and emission mechanisms of the various types of magnetar burst are widely debated \citep[see][for a recent review]{Turolla2015a}. However in the case of the giant flares, energetics considerations motivate a consensus that an optically thick pair-plasma fireball is formed, which is trapped in the magnetic field \citep{Thompson1995, Thompson2001a, Heyl2005a}. The energy for the burst perhaps comes from some form of magnetic reconnection event, caused by the gradual decay of the magnetic field. Part of the energy output is emitted in the initial spike of radiation, while part of the burst energy forms a pair plasma through rapid pair creation, and this plasma is trapped in the closed magnetic field lines of the magnetar in the form of a fireball. This fireball then gradually evaporates through radiative cooling, putatively causing the observed slowly decaying tail of the light curve. This model has been used very successfully to fit the phase-averaged decaying tail of the light curve \citep[see][]{Feroci2001a, Hurley2005a}.  The fireball model may also be applicable to intermediate flares, as the creation of a fireball is unavoidable when a sufficiently large amount of energy is injected into the magnetosphere \citep[see][]{Thompson1995}, and at least some intermediate flares also have long tails and pulsations \citep[see][for a discussion of the morphologies of intermediate flares]{Turolla2015a}. For short bursts, the applicability of the fireball model is debatable, as these bursts usually do not have an observable decaying tail.

The presence of a localized emission region (in the form of a trapped fireball) is not sufficient to explain the periodicity in the light curves, as while one or more emission regions rotating into and out of the line of sight would create rotational modulations to the light curve, these modulations would change dramatically as this emission region shrinks.  This is not what is observed, as while the shape of the observed modulations can change dramatically during the tail of the light curve, it usually stays constant for large parts of the tail \citep[see][]{Feroci2001a, Ibrahim2001a, Hurley2005a}.  This suggests the modulations are caused by some form of beaming, rather than by the physical restriction of the emitting region to hot spots. It also suggests that the shape of the pulse profile is not determined by the size of the fireball, but rather by another physical characteristic of the system, and that this characteristic can change during the flare, but does not always change. 

In discussing the beaming of photons in a magnetar environment we have to distinguish between two photon linear polarization normal modes that behave differently in a magnetar-strength magnetic field. At low frequencies, these modes have very different cross sections \citep[see][]{Herold1979} for Compton scattering off free electrons, which is the dominant source of opacity in a magnetar atmosphere \citep{Thompson1995}. Photons in the ordinary mode (O-mode) are polarized such that their electric field vector lies in the plane formed by their momentum and the magnetic field, and have scattering cross section roughly equal to the Thomson cross section, while photons in the extraordinary mode (E-mode) are polarized perpendicular to this plane, which strongly inhibits scattering. As such, well below the cyclotron frequency, E-mode photons have a scattering cross section much lower than the Thomson cross section, which scales approximately as the radiation frequency squared divided by the magnetic field strength squared. This means that close to the magnetar, where the magnetic field is strongest, E-mode photons may diffuse freely while O-mode photons couple strongly to the matter.  The result of this is that in regions of high magnetic field strength, the radiation flux will be dominated by photons in the E-mode. However, O-mode photons do provide a significant contribution to the radiation force \citep[see][]{Miller1995,van-Putten2013a}.  Note that these two modes are convenient choices for modelling scattering.  They are approximate polarization eigenstates of a photon propagating in a uniform magnetic field in either the vacuum or in a plasma, and yield a fully correct probabilistic description of radiative transfer in neutron star magnetospheres.

Because the O-mode photons couple strongly to matter through Compton scattering they can easily be beamed by some form of relativistic outflow. \citet{Thompson1995} note that radiation escaping from the base of the fireball can drive such a relativistic outflow, as the initial spike of the flare will ablate material off the surface of the neutron star, which can then be accelerated by the super-Eddington luminosity in the high-opacity O-mode. O-mode photons advected with this flow will then get beamed along the direction of the flow. However, since any outflow has to follow the magnetic field lines, which diverge from each other, this beaming will not be strongly peaked in a single direction. The phase width of this outflow zone radiation will be a direct measure of the range of colatitudes spanned by the open field lines at the altitude that the flow becomes optically thin for the O-mode photons.  The low-opacity E-mode photons would not get beamed this way. 

\citet{Thompson2001a} refine this beaming scenario, providing a way in which both photon modes can get beamed. While the E-mode X-ray photons have very low opacity close to the star, this opacity increases rapidly away from the star (for photon frequencies far below the cyclotron frequency it scales approximately with the  inverse square of the magnetic field strength).  When the altitude is such that the cyclotron frequency is close to the E-mode photon frequency, its opacity is similar to that for the O-mode.  In addition, there is a degree of mode-switching in scattering events that enhances the overall opacity of the E-mode.  If there is matter suspended higher up in the magnetic field of the star, supported against gravity by the high luminosity, but unable to escape because it is on closed field-lines, this forms an optically thick barrier to E-mode photons. The way for the E-mode photons to escape is to push this matter aside, creating a sort of nozzle.  This nozzle then causes beaming of the E-mode photons, while the O-mode photons still get beamed by advecting along a relativistic outflow.

Up to now the fireball model has only been used to fit phase-averaged light curves of giant flares,  showing that the energy released by a shrinking fireball is an excellent fit for the observed light curves \citep{Feroci2001a,Hurley2005a}. The goal of this paper is to start using the fireball scenario to also model the pulsations in those light curves, and make the beaming predictions of \citet{Thompson2001a} quantitative, by making the simplest possible fireball model that produces the desired beaming.  Additionally, we want to start to systematically discern the character of the immediate environment around a bursting magnetar.  

We set up a model of a fireball trapped in a dipolar magnetic field that emits radiation close to the surface of the star. Outside the fireball we create a relativistic outflow, simulating the matter ablated from the surface of the star.  This setup is effectively the canonical trapped evaporating fireball scenario for the tails of magnetar giant flares, as proposed in the classic papers of \citet{Thompson1995,Thompson2001a}.  It assumes that the arguments put forward in those papers for the formation of both the fireball and the associated outflows, driven by the super-Eddington luminosities, are valid.   Note that in our model the outflow is imposed {\it a priori}, so the radiation field and the outflow in our model are not fully self-consistent (see Section \ref{sec:outflow} for details).  We then use a Monte Carlo radiation transfer code to scatter photons through this setup, and track the direction in which the photons escape the system\footnote{A similar strategy has been used to compute quiescent magnetar spectra in the twisted magnetosphere model, \citep[see for example][]{Fernandez2007,Nobili2008a,Nobili2008b}. In the quiescent case the scattering electrons are in currents, whereas in the giant flare case they are in the outflow.}. By varying the physical parameters of our model, particularly the size of the fireball, magnetic field strength and density and velocity of the outflow, we test how these parameters influence the degree to which the radiation becomes beamed. The fireball size is of particular interest, as observations show the pulse profile should be largely independent of this parameter. We also perform some simulations with slightly more complicated geometrical setups, to recreate more closely the beaming scenario set out by \citet{Thompson2001a}. Our simulations include special relativistic beaming effects as well as general relativistic light bending (important close to the star), take the vacuum and cyclotron resonances into account, and allow photons to convert between the two polarization modes where appropriate.

In addition to tracking the angular intensity profile of the escaping radiation, our radiative transfer simulations also give us spectral and polarization information. This is not a realistic output spectrum, as we do not include any sources of opacity other than electron scattering.  However, we can track differences in the spectrum for different angular directions, to test whether we can recreate the spectral variations with rotation found in some magnetar flares \citep{Feroci2001a, Boggs2007a}. We also test whether the pulse profile is significantly different between the two polarization modes, something which might be observable with proposed X-ray polarimetry missions \citep[see for example][]{Fernandez2011a}, such as {\it XIPE} \citep{Soffitta2013a}, {\it IXPE} \citep{Weisskopf2014a}, {\it XTP} \citep{Dong2014a} and {\it PRAXyS} \citep{Jahoda2015a}. Such information offers the prospect for better understanding the geometrical relationship between the outflow, nozzle and fireball zones in future models that are more self-consistent.

The layout of this paper is as follows. In Section \ref{sec:model} we set out the geometrical and physical setup of our model, detailing the assumption and approximations we make with regards to the properties of the fireball and the outflow in our model, and discuss the scattering opacity. In Section \ref{sec:method} we discuss our numerical method, describing in detail how a photon propagates through our model. In Section \ref{sec:results} we present the results of our simple geometrical model, and discuss how these depend on the various physical parameters. In Section \ref{sec:complicated} we make some additions to our geometrical model to more closely resemble to scenario set out by \citet{Thompson2001a}, and discuss the results of these models. Finally in Section \ref{sec:conclusion} we summarize our results, foremost among which is the necessity of a collimated relativistic outflow in order to account for the observed pulse fraction.  Therein we discuss what they mean both for past and future observations.

\section{Model setup}
\label{sec:model}

In the fireball model for magnetar flares \citep{Thompson1995}, a volume of hot pair plasma is trapped in the magnetic field of a magnetar, and gradually shrinks as it radiates away its energy. The fireball is expected to have an initial volume  of the same order of magnitude as  the volume of the neutron star \citep{Thompson2001a}, and gradually shrink until it evaporates completely. Outside this fireball, matter will be ablated off the surface of the neutron star, and this matter will form an outflow due to the highly super-Eddington luminosity radiated by the fireball \citep{Thompson2001a}. 
 
The shape of the fireball is an open question. \citet{Thompson2001a} considered spherical or cylindrical structures, but the exact shape trapped in a realistic magnetar field structure, which may sustain local twists, is unclear.  \citet{Feroci2001a} showed that the time-dependence of the X-ray luminosity $L_x(t)$ for the light curve of the giant flare from SGR 1900+14 on 27 August, 1998 was well fit by the following function:
\begin{equation}
   L_x(t) = L_x(0)\left(1 - \frac{t}{\tau_\mathrm{evap}}\right)^\chi
\end{equation}
where $\tau_\mathrm{evap}$ is an evaporation timescale.  The parameter $\chi = a/(1-a)$ can be related, in a simple model where the fireball has uniform energy density and surface flux, to the shape of the fireball \citep{Thompson2001a}.  For the 1998 giant flare from SGR 1900+14, the best fit value was  $a=0.75$ \citep{Feroci2001a}, while for the 27 December, 2004 giant flare of SGR 1806-20, \citet{Hurley2005a} found $a=0.6$.  A homogeneous fireball cannot have a value higher than $a=2/3$, which is also the value for a homogeneous sphere, so these results show the SGR 1900+14 fireball may have a more complex structure \citep{Thompson2001a}.  In this paper we choose to model the magnetic field as a pure dipole, which is sufficient to explore the generic character of outflows and fireballs, and their coupling.  This choice leads to a torus-shaped fireball, as that is the only shape that can be trapped in the closed field-lines of a dipole field.   We use the equations for a dipolar magnetic field in full general relativity given by \citet{Wasserman1983a}. When considered in two dimensions (as our model geometry has rotational symmetry) the boundary of the fireball lies along a single closed field line, which is determined by a single parameter. We choose to use the diameter of the torus-shaped fireball as this parameter, defined as the equatorial radius of the outermost field line minus the radius of the star.  We let the diameter vary between one and one hundred kilometres ($\sim$ 0.1-10 stellar radii), to cover all possible fireball sizes. 

\begin{figure}
\begin{center}
\includegraphics[width=\columnwidth]{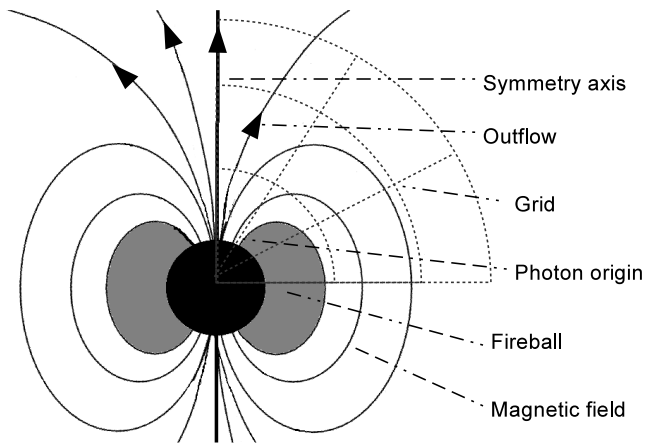}
\end{center}
\caption{Schematic of the geometry of our simulations. Photons propagate from their point of origin until they are destroyed by hitting the star or the fireball, or until they escape at the top of the grid. The outflow occurs everywhere outside the fireball. This image is not to scale, as in our models the neutron star radius is $10^6$ cm, while the top of the grid is located at $r=3\times 10^7$ cm, which is the point where we let photons escape.
}
\label{fig:geometry}
\end{figure}

The radiation is expected to escape into the outflow zone primarily at the base of the fireball, where its scattering opacity is lowest due to its dependence on magnetic field strength. For simplicity, we let all radiation originate here. This is a rough approximation, but we find that the properties of the escaping photon field are mostly determined higher up in the outflow, so the exact starting point of the radiation is not very important. A schematic of our model setup can be seen in Figure \ref{fig:geometry}. 

In the case of a giant flare, the fireball is expected to have a temperature on the order of 0.1-1 MeV \citep[see also the discussion in][]{Harding2006}. However, the typical energy of the escaping photons is much less than this, as the best fit blackbody temperature is typically around 10 keV, although it is typically hotter in the initial spike of such a flare. Small burst spectra tend to have similar best-fit blackbody temperatures. This large disparity in ``injected'' and emergent energies is mostly caused by photon splitting \citep[see][]{Thompson1995,Baring1998a}, whereby a single photon spontaneously converts into two photons of roughly half the energy in the presence of a very strong magnetic field. The observed temperatures of around 10 keV also coincide well with the temperature at which photon splitting is expected to freeze out, which is given by \citet{Thompson2001a} as a blackbody temperature of 11 keV \citep[see also][]{Baring1997}. The photon field will reach this temperature at the splitting photosphere, which is approximately the point where the magnetic field strength equals $B_{\rm cr}=4.4\times10^{13}\,{\rm G}$, the value for which the cyclotron energy of an electron is equal to its rest mass. For typical surface magnetic field strengths of $10^{14}-10^{15}$ G this point is reached at a height of a few neutron star radii.

In our outflowing atmosphere models the point of largest optical depth falls around the cyclotron resonance, the point where the photon frequency is equal to the electron cyclotron energy. This occurs higher up in the atmosphere than the photon splitting photosphere, which means that we expect all photons to be scattered multiple times after their last splitting event. Because of this, and for the sake of simplicity, we choose not to include photon splitting in our models, presuming that it mainly establishes the range of photon energies injected into the outflow zone. Instead, we only consider opacity from Compton scattering off free electrons (see Section \ref{sec:scattering}), which is the dominant source of opacity in a magnetar atmosphere \citep{Thompson1995}.  We take our input spectrum as a blackbody with a temperature of 10 keV,  the rough spectrum expected outside the photon splitting photosphere \citep[see][for more accurate spectral models]{Ozel2001a, Lyubarsky2002a}. We also run some models with different blackbody temperature to test the effects of higher or lower photon energies, to explore the impact of raising or lowering the photon energy, which should alter the net E-mode opacity.  Note that since we focus on the beaming of the radiation, computing an accurate spectrum is outside the scope of this paper. We do look at spectral variations in our results, but only consider changes in the average energy of the spectrum.  As such, the precise shape of the input spectrum is of secondary importance.   We also varied the polarization fraction of the injected spectrum (photon splitting should generate a preponderance of O-mode photons in the injection), but found that our results were insensitive to this parameter.

\subsection{Outflow properties}
\label{sec:outflow}

We fix the properties of the outflow in our model in advance, as computing these properties from the assumed radiation field in a self-consistent manner is outside the scope of this work, and we want to keep our model as simple as possible. For example, significant acceleration due to Compton drag precipitated by the intense radiation bath will impact the outflow speed and its radial profile.  Notwithstanding, we do base the density in the outflow on theoretical estimates of the amount of mass lost through this outflow, which we will discuss below. For the velocity of the outflow we assume a simple power law: 
\begin{equation}
   v(r) = v_{\rm start} + (v_\infty-v_{\rm start})\left(1-\frac{R_\star}{r}\right)^\beta,
 \label{eq:v}
\end{equation}
where $v_{\rm start}$ is the outflow velocity at the surface of the star, $v_\infty$ is the outflow velocity at infinity, $R_\star$ is the radius of the neutron star and $\beta$ is the power-law index. This choice of a beta-law velocity profile is arbitrary, as this work is intended to be exploratory, and there are no magnetar outflow models in the literature. We choose to use a beta-law profile as this is a common choice in stellar wind theory. Moreover, for $v_{\infty}>v_{\rm start}$, it embodies the general acceleration property $dv/dr > 0$.  We set $v_{\rm start}=0$ in all our models, as  there can be no outflow velocity through the solid crust of the star, and compute models for different values of $v_\infty$ in the range $0.1c$ to $0.95c$. The ranges over which we vary the various parameters in computing our results can also be seen in Table \ref{tab:parspace}.  For $\beta$ we use a value of $0.8$, which is a fairly arbitrary choice from the typical range this parameter takes in winds from massive stars. In fact, we find that using a different value of $\beta$ has a similar effect as using a different value for $v_\infty$, so we choose not to vary this parameter in our results. The velocity of our outflow is always directed along the local direction of the magnetic field, since charged particles will follow the field lines, and the high surface temperature will ensure the ablated matter is fully ionized, if it is primarily composed of hydrogen, helium or light elements.

The density of our outflow is determined by mass continuity, and can thus be computed everywhere from a single constant value of the mass loss rate. We will now estimate this mass loss rate in two different ways. \citet{Thompson1995} compute the total amount of mass ablated from the surface of the neutron star by a burst as $\Delta M \simeq E_{\rm th} / gR_\star$, where $E_{\rm th}$ is the total amount of heat absorbed by the crust, and $g$ is the gravitational acceleration on the surface of the star. They also estimate that in a short burst $E_{\rm th}$ is approximately $10^{38}$ erg, which leads to a total mass loss of $\sim 5\times 10^{17}$ g for typical neutron star parameters. If we scale this to a giant flare, which produces on the order of $10^4$ times as much energy, and divide by a duration of $\sim500$ s, we find a mass loss rate of approximately $10^{18}$ g s$^{-1}$. Our second estimate of the mass loss rate is based on \citet{Thompson2001a}, who estimate the mass loss rate of the outflow in their calculations as $\dot{M} c^2 \lesssim L_{\rm Edd} \left(\frac{GM}{R_\star c^2}\right)^{-1}$, where $L_{\rm Edd}$ is the classical Eddington luminosity of $1.8\times10^{38}$ erg s$^{-1}$ (for a 1.4 solar mass neutron star with a hydrogen atmosphere). For typical neutron star parameters, this gives a mass loss rate of approximately $10^{17}$ g s$^{-1}$.  The roughly commensurate nature of these two estimates should not be over-interpreted.  Electron-positron pair creation is expected to be rife in the fireball and outflow, so that the effective Eddington luminosity drops by the $m_e/m_p$ mass ratio.  Nevertheless, these estimates serve in our model to roughly benchmark the boundary conditions at the base of the outflow.

As the matter follows the magnetic field lines, which curve outwards, the density in our outflow falls off more quickly than it does in the typical mass continuity case for a spherical flow, for which solid angles in the wind are conserved. Since the dipolar field structure couples radius $r$ and colatitude $\theta$ via $r\propto \sin^2\theta$, the solid angle subtended by the outflow at the star's centre should scale roughly as $\theta^2 \propto r$ at low altitudes near the dipole axis.  This implies a mass dilution factor $\propto r^{-3}$ is appropriate for a dipolar field morphology at polar colatitudes.  This means the density is given by $\rho = \rho_0 \times v_0/v \times(r_0/r)^3$. We use the subscript 0 to refer to values in our bottom grid cells (see Section \ref{sec:method}) rather than right at the surface of the star, as right at the surface of the star the velocity has to be zero (due to the solid crust) so mass continuity would give an infinitely large density there. The velocity $v_0$ is simply $v$ as given by Equation \ref{eq:v}, with $r$ the midpoint of the bottom grid cells, while $\rho_0$ is an input parameter. For our baseline model with $v_\infty=0.2c$ we set $\rho_0=0.01$ g cm$^{-3}$, which when taking the solid angle of the outflow into account gives a mass loss rate of approximately $10^{18}$ g s$^{-1}$ (varying somewhat with the size of the fireball). We also compute results for different values of $\rho_0$, and when computing results for different values of $v_\infty$ we change $\rho_0$ accordingly to keep the same mass loss rate.

As the outflow makes up only a small fraction of the total energy loss rate of the flare, we expect the temperature of the outflowing matter to follow the temperature of the photons, rather than the other way around. Thus, we choose not to give the outflow a temperature, but rather to let all scatterings conserve photon energy in the frame comoving with the flow. While this does not give the correct shape for the Comptonized output spectrum, it does give the right average photon energy, as in a flow dominated by the photon field the small matter component should not be able to change the average energy of the photons. Note that we compute this conservation in the reference frame comoving with the outflow. The average photon energy in the observer frame does change due to special and general relativistic effects.  This means we also ignore any beaming effects from relativistic random thermal motion of the electrons. We expect these effects to be minor, as the typical blackbody temperature of a magnetar flare spectrum of approximately 10 keV is only a small fraction of the electron rest mass energy of 511 keV.

\subsection{Compton scattering in strong magnetic fields}
 \label{sec:scattering}

In a very strong magnetic field, the Compton scattering cross section is very different from the non-magnetic case \citep{CLR71,Herold1979}, exhibiting many resonances at the cyclotron frequency $\omegaB =eB/m_{\rm e}c$ and its harmonics \citep{DH86}.  Here, $e$ and $m_{\rm e}$ are the charge and mass of an electron, respectively and $B$ is the magnetic field strength. For the model construction here, near the stellar surface, the highly supercritical field guarantees that the X-rays mostly sample domains around or below the cyclotron fundamental.  In such parameter regimes, there can be a vast disparity in the scattering cross section for the two linear polarization modes.  Well below $\omegaB$, the O-mode photons have a scattering cross section similar to the classical Thomson value, $\sigt$, while E-mode photons have their cross section strongly reduced below $\sigt$. The cross section for E-mode photons can be approximated as $\omega^2/\omegaB^2$ times the Thomson cross section, where $\omega$ is the photon frequency.  In the cyclotron resonance, the cross sections for the two modes are comparable, but not equal.  Note that for photons travelling close to parallel to the magnetic field the distinction between the E-mode and the O-mode fades, with circular polarization serving as the appropriate mode description.   Then all photons have reduced cross sections as long as $\omega \ll \omegaB$: see \citet{Herold1979} for a description of scattering formalism employing both circular and linear polarization states.

In the magnetic Thomson domain, a classical formalism for polarization eigenmodes and scattering in plasma was presented in \citet{CLR71}.  This inherently embeds information on magnetized plasma dispersion, such as is appropriate for analyzing Faraday rotation.  In the quantum domain, one needs to also incorporate the influence of the magnetic field, namely vacuum birefringence.  A principal consequence is that the O-mode and E-mode photons propagate with slightly different speeds, and so their electric field vectors rotate about {\bf B}, a vacuum polarization analogue of Faraday rotation.  For our problem, we need to include both these dispersive elements, and accordingly we use the opacity given by \citet{Ho2003a}.  In such a unified description, the contribution to the refractive index $n$ for vacuum polarization scales as $n-1 \propto \fsc\, (B/B_{\rm cr})^2$, where $B_{\rm cr}=m_{\rm e}^2c^3/(e\hbar) \approx 4.41\times 10^{13}\,$Gauss is the quantum critical field, at which the cyclotron energy equals the electron rest mass energy, and $\fsc = e^2/\hbar c$ is the fine structure constant.  The contribution from plasma dispersion scales the refractive index via $n-1\propto -\, (\omega_p/\omega)^2$, where $\omega_p = \sqrt{4\pi n_{\rm e}e^2/m_{\rm e} }$ is the plasma frequency.  The competition between the frequency-independent vacuum dispersion and the frequency-dependent plasma contribution establishes a resonance feature, commonly termed the {\it vacuum resonance}, at a photon frequency $\omega_{\rm res}$ that depends on both the plasma density and the magnetic field strength: $\omega_{\rm res}\propto \omega_p\, \sqrt{\fsc}\, (B/B_{\rm cr})$.   For magnetar atmospheres, the vacuum resonance typically arises in the soft X-ray window.  See \citet{Harding2006} for a review of this hybrid picture of dispersion for neutron star magnetospheres.

For linearly-polarized photons that Thomson scatter from mode $j$ to mode $i$ ($i,j=1,2$ for E-mode, O-mode), a compact expression for the opacity in strong magnetic fields is given by:
\begin{equation}
   \kappa_{ji} \; =\; \frac{n_{\rm e}\sigt}{\rho} \sum_{\alpha=-1}^1 \frac{\omega^2}{(\omega+\alpha\omegaB)^2 
   + \Gamma_{\rm e}^2/4} \left|e_\alpha^j \right|^2 A_\alpha^i \quad .
 \label{eq:kappa}
\end{equation}
In this equation $n_{\rm e}$ is the electron number density, $\sigt$ is the classical Thomson scattering cross section, and 
in $B\ll B_{\rm cr}$ domains, $\Gamma_{\rm e}= 2e^2\omegaB^2 /(3m_{\rm e}c^3)$ is the linewidth of the cyclotron resonance.  
The vector $\boldsymbol{e}_j$ is the normal mode polarization vector, whose components are given by:
\begin{align}
   \left| e_{\pm 1}^{\,j} \right|^2  & \; =\; \frac{(1 \pm K_j \,{\rm cos}\,\theta )^2}{2(1+K_j^2)}  \nonumber \\
   \left| e_0^{\,j} \right|^2  & \; =\;  \frac{K_j^2 \,{\rm sin}^2\,\theta}{1+K_j^2}\quad ,
\end{align}
where $\theta$ is the angle between the incident photon and the magnetic field vector. $K_j$ is a term that incorporates the influences of vacuum and plasma dispersion. When both dispersion effects are ignored this term reduces to zero for the E-mode and to infinity for the O-mode, reducing the opacity to the form given by \citet{Herold1979}.  When plasma dispersion is included, but vacuum polarization influences are neglected, the formalism maps over to that offered in \citet{CLR71}.  For the precise form of $K_j$ the reader is referred to \citet{Ho2003a}.  Finally, $A_\alpha^i$ is the angle integral given by
\begin{equation}
   A_\alpha^i \; =\; \frac{3}{4} \int \left|e_\alpha^i (\theta') \right|^2 {\rm sin}\,\theta' \,{\rm d}\theta' \quad ,
\end{equation}
where $\theta'$ is the angle between the scattered photon and the magnetic field vector.

Equation (\ref{eq:kappa}) describes a photon scattering from polarization mode $j$ and angle with respect to the magnetic field $\theta$ to polarization mode $i$ and angle $\theta'$. The distribution of post-scattering angles is thus contained in $A_\alpha^i$. We can construct a probability density function for the post-scattering angle by simply differentiating Equation (\ref{eq:kappa}) with respect to $\theta'$. We will discuss how we draw a random post-scattering direction from this distribution function in Section \ref{sec:angleselect}.

In the rest frame of a cold plasma, for which kinetic motions of electrons are negligible,
the opacity for a photon in polarization mode $j$ and incident angle $\theta$ to scatter to any polarization mode and any angle is given by:
\begin{equation}
   \kappa_{j} \; =\; \frac{n_{\rm e}\sigt}{\rho} \sum_{\alpha=-1}^1 
   \frac{\omega^2}{(\omega+\alpha\omegaB)^2 + \Gamma_{\rm e}^2/4} \left| e_\alpha^j \right|^2 A_\alpha \quad ,
\label{eq:kappa_j}
\end{equation}
with $A_\alpha = \sum_{i=1}^2 A_\alpha^i$. It can be shown that when employing transverse modes, as is done here, $A_\alpha=1$ \citep{Ho2003a}, greatly simplifying the form of this opacity. This is the opacity we use to compute the path length of a photon in our simulations.

The opacity formulation here is not a fully general description for scattering in superstrong magnetic fields.  It is precise for highly sub-critical fields, $B\ll B_{\rm cr}$.  When the field rises to near-critical strengths and even higher, Eq.~(\ref{eq:kappa}) can be accurately applied for $\hbar\omega\lll m_ec^2$, i.e. low frequency photons with $\omega \ll \omegaB$.  If the photons are in the hard X-ray or gamma-ray bands, Klein-Nishina and recoil effects become significant and reduce the value of the cross section: this arises to some extent in the rest frame of the outflow due to the blueshifting of photon energies.  Moreover, above the cyclotron fundamental, a multitude of harmonic resonances appear \citep{DH86}, and these are not captured in the form in Eq.~(\ref{eq:kappa}).  These are complications that are beyond the scope of our present work.  We believe that our choice for the opacity captures the general character for much of the phase space applicable to our magnetar outflow problem.  This contention is underpinned by the fact that the radiative transfer in the outer extremities of the outflow occurs in sub-critical fields where Eq.~(\ref{eq:kappa}) is applicable.

Notwithstanding, in the cyclotron resonance and its harmonics, the simple ``non-relativistic'' linewidth $\Gamma_{\rm e}= 2e^2\omegaB^2 /(3m_{\rm e}c^3)$ discussed for Eq.~(\ref{eq:kappa}) is inaccurate, overestimating the width by orders of magnitude when $B\gg B_{\rm cr}$.  This common invocation would therefore yield erroneous estimates of the opacity in and near the cyclotron fundamental, and so must be upgraded to treat decay rates for general magnetic field regimes \citep{Latal86,BGH05}.  Since results are readily available for the decay widths $\Gamma_{\rm e}$ in $B\gtrsim 0.1 B_{\rm cr}$ fields that are amenable for implementation in numerical codes, this high-B refinement is accurately captured in our simulation --- fully relativistic cyclotron decay widths $\Gamma_{\rm e}$ pertinent to the cyclotron fundamental are employed in this paper, the details of which are described in the Appendix.  Note that the interplay of spin-dependent influences in the Compton cross section evaluation in the resonances \citep{Gonthier14} will be neglected here, since they will have only a small influence on the character of the results presented below.

\section{Numerical method}
\label{sec:method}

Our method consists of propagating photons through a spatial grid one by one until they escape at the top of the system or are destroyed. We describe the properties of the spatial grid here, detailing the propagation of a photon in Section \ref{sec:propagation}, and our method for selecting a scattering angle in Section \ref{sec:angleselect}. A schematic of the geometry of our model can be seen in Figure \ref{fig:geometry}.

Because the fireball is a torus in our model and the magnetic field has no azimuthal component, our system is completely rotationally symmetric. We construct a two-dimensional grid of cells of fixed size in radial and polar coordinates, with the grid running from the surface of the neutron star at $r=10^6$ cm to a fixed end point at $r=3\times 10^7$ cm in the radial direction, and from $\theta=0^\circ$ to $\theta=90^\circ$ in the polar direction. We divide this grid in 300 cells in the radial direction and 100 cells in the polar direction, and determine the properties of each cell at the midpoint of that cell. Testing shows that  increasing the number of grid cells does not change our results. 

The top of our grid is chosen to fall well above the photosphere. The altitude at which the cyclotron resonance coincides with the spectral window of $1-30$ keV typically occurs slightly below $r=10^7$ cm, and we would expect the outflow to be fairly optically thin above that. We find that between $r=10^7$ cm and $r=3\times 10^7$ cm photons scatter slightly less than once on average, which gives us confidence that the top of our grid is sufficiently high. In the polar direction, we remove all grid cells whose midpoint falls inside the fireball. Any photons that travel out of the grid in this direction, or into the surface of the star, are destroyed, as if completely absorbed in the fireball. Photons exiting the grid at $\theta=0$ are simply mirrored back into the same cell, as this is the symmetry axis of our system, while only photons escaping at the top of the grid are added to our output results.

While the physical properties of our fireball and outflow system can be described fully in two dimensions, the movement of the photons in the azimuthal direction is relevant, as the azimuthal angle partially determines the angle between the photon and the local magnetic field direction, which is an important factor in the scattering cross section. Thus, the azimuthal components of the photon momenta are recorded at all positions, with any movements and coordinate transformations being carried out as in a three dimensional system, but without keeping track of the azimuthal position of the photon.

Our computational method consists of propagating a large number of photons from the point where we create them until they either escape at the top of the grid or are destroyed.  When a photon escapes at the top of the grid, we add a single photon to the correct bin in a three-dimensional array of different escape angles, photon energies and photon polarization modes.  We run our code for a fixed amount of time rather than for a fixed number of photons, as we find that both the number of photons that can be propagated in a certain amount of time and the fraction of photons that escape vary by several orders of magnitude depending on the physical parameters. The results shown in this paper have been simulated in approximately 200 hours of CPU time each,  on a 2.6 GHz Intel Xeon CPU (with the exception of Figures \ref{fig:spectrum_density} and \ref{fig:spectrum_angle}, which are based on 4000 and 1000 hours of CPU time, respectively).

In the propagation of a single photon, we take into account both general and special relativistic shifts in the direction and frequency of the photon.  Doing this, we also naturally include the advection of photons with the flow in regions of high optical depth. This is especially relevant for photons near the cyclotron resonance, for which advection is the main mode of moving outwards. Below the cyclotron resonance, E-mode photons generally diffuse quite freely, while O-mode photons move either by advection or by converting to the E-mode.  This advection enters our model through the bias along the direction of the flow that is introduced when transforming the photon direction from the comoving frame to the stationary frame. We have carried out tests that confirm that photons in a region of large optical depth on average advect outward with the flow at a rate approximately equal to the outflow velocity.

We have tested the basic soundness of our scattering method  and our code by recreating the results of \citet{Miller1995}, specifically the fraction of escaping photons in a given mode for monochromatic photons propagating through a homogeneous atmosphere for different photon and cyclotron frequencies. We have also used the code to compute the flux, radiation force and polarization mode distribution throughout the static atmosphere solutions of our previous paper \citep{van-Putten2013a}, and comparing those results to the integral based computations we performed in that work.

The Doppler boosting of photons to the observer's frame does not impact the polarization configuration.  The opacity determinations are made in the frame of the cold outflow. In general, when Lorentz boosting to the observer's frame, the electric field vector of a photon is tilted \citep[an issue that is much discussed in the gamma-ray burst and blazar jet literature, see for example][]{Lyutikov2003}.  However, in these jet contexts, the boost direction is typically not aligned with the magnetic field.  The magnetar problem is inherently different since, in the absence of significant rotation, the plasma flows along field lines (note that this does not hold true when one gets close to the light cylinder, but we never realize that regime in our study).  Lorentz boosts along the magnetic field lines preserve both the strength of the local neutron star magnetic field and the azimuthal direction (or phase) of any electric field present about the boost direction.  Photons in our problem are not all travelling along the local magnetic field (the boost direction).  However since we characterize polarization using the standard E/O mode classification, their electric field vectors lie either in the {\bf k}-{\bf B} plane (at one particular phase around {\bf B}), or orthogonal to this ($\pi$/2 removed from the former phase).  Looking down the field line {\bf B}, a boost rotates neither of these electric field vectors, leaving their phases unchanged, and will only change the {\bf E}-field magnitude parallel to and perpendicular to {\bf B}, coupling to the aberration formula.  The invariance of linear polarization state under Lorentz boosts along {\bf B} is also discussed, for example, in Appendix A of \citet{Beloborodov2013}.  Thus by formulating the problem using linear polarization modes rather than elliptical polarizations, there is no rotation in the plane of polarization as a result of Lorentz boosts.

\subsection{Photon propagation}
\label{sec:propagation}

The step by step by step propagation of a single photon in our model proceeds as follows:
\begin{enumerate}
\item The photon is created at the midpoint of the grid cell closest to the base of the fireball. It is given a random direction, with the constraint that its initial direction is away from both the neutron star and the fireball. For the initial energy of the photon we iterate repeatedly through a pre-computed list of one hundred equal probability photon energies, computed by integrating over a normalized blackbody function and saving the energies at which this integral is an integer multiple of  $10^{-2}$.

\item The photon is assigned a random travel distance in optical depth units of $\Delta \tau=-{\rm ln}\,x$, with $x$ a uniform random number between zero and one. This distance is then converted to a travel distance in physical units $\Delta r = \Delta \tau / (\kappa_j \rho)$, where $\kappa_j$ is the optical depth for a photon to scatter to any outgoing mode and angle, as given by Equation \ref{eq:kappa_j}, and $\rho$ is the density of the cell.

\item The photon propagates. If the photon comes to a boundary of the cell before completing its transit over $\Delta r$, the remaining distance is converted back to optical depth units, and then to physical units in the new cell, which has different density and opacity. The propagation is then continued and the sequence if repeated until the optical depth is fulfilled and a scattering event ensues.

\item Whenever a photon changes cell, we check whether it crossed the vacuum resonance in moving from one cell to another. This resonance occurs at photon energy \citep{Ho2003a}:
\begin{equation}
E_{\rm vac} \simeq 1.02\left(\frac{Y_{\rm e}\rho}{1\,{\rm g cm}^{-3}}\right)^{1/2} \left(\frac{B}{10^{14}\,{\rm G}}\right)^{-1} f(B) \, {\rm keV}, 
\end{equation}
where $Y_{\rm e} = Z/A$ is the electron fraction and $f(B)$ is a slowly varying function of order a few \citep[see][for details]{Ho2003a}. If a photon crossed this resonance it has a chance of adiabatically converting to the other polarization mode. This chance is given by $P_{\rm con} = {\rm exp}[-0.5\pi (E/E_{\rm ad})^3]$, with $E$ the photon energy and $E_{\rm ad}$ given by:
\begin{equation}
E_{\rm ad} = 2.52[f(B)\,{\rm tan}\,\theta]^{2/3} \left|1-\frac{\omega_{\rm C,ion}^2}{\omega^2}\right|^{2/3}\left(\frac{H_\rho}{1\,{\rm cm}}\right)^{-1/3},
\end{equation}
where $\theta$ is the angle between the photon and the magnetic field, $\omega_{\rm C,ion}$ is the ion cyclotron frequency and $H_\rho$ is the density scaleheight along the ray. We use this condition to let a photon convert polarization modes when appropriate.

\item The electron cyclotron resonance also receives special treatment, as it is much narrower than the size of the cells, but cannot be treated as infinitesimally thin. To prevent photons from propagating past the cyclotron resonance in one step, and to ensure they interact with the resonance in a realistic manner, we make sure that when a photon is near this resonance it only propagates in small steps and its opacity is updated after every propagation (rather than only when it scatters or enters a new cell). This is done through a check that occurs when a photon enters a new cell, assessing whether the magnetic field strength at which the photon's frequency equals the cyclotron frequency falls in between the minimum and maximum magnetic field strength of the new cell. This same check also occurs when a photon scatters to a new frequency. If the cyclotron resonance for the current photon does indeed fall inside the current cell, the photon is prevented from moving by more than one tenth of the approximate geometrical extent\footnote{The width of the resonance is in principle a relative width in frequency units. This can be converted to a width in $B$ through the scaling of the cyclotron frequency with $B$, which can then be converted to length units through the scaling of $B \propto r^{-3}$} of the cyclotron resonance per step, and its opacity is updated after every step. This updated opacity is calculated from the magnetic field strength at the photon's accurate position, rather than from the pre-calculated magnetic field strength at the centre of the cell, as is done when a photon is not near the cyclotron resonance. This restriction on the maximum travel distance of the photon per step is lifted when it enters a new cell. 

\item At the new position of the photon, we convert its direction from the reference frame of a global observer to that of a local stationary observer, to incorporate general relativistic beaming effects. We use the definition of the dot product in General Relativity,

\begin{equation}
\mu_{\gamma s} |p| |dx| = p^\alpha dx^\beta (g_{\alpha \beta} + s_\alpha s_\beta )
\end{equation}
where $\mu_\gamma$ is the cosine of the polar direction of the photon and $\mu_{\gamma{\rm s}}$ the same cosine in the local stationary reference frame (the azimuthal direction of the photon is not affected by this frame shift), $p$ is the four-momentum of the photon, $dx$ the $x$-axis (0,1,0,0), $s$ a local stationary observer, $s^\alpha = ((1-R_g/r)^{-1/2}, 0, 0, 0)$ and $g$ is the Schwarzschild metric, we obtain
\begin{equation}
   \mu_{\gamma{\rm s}} \; =\;  \mu_\gamma \left(1-\frac{R_{\rm g}}{r} \right)^{-1/2} \left[ 1+ \mu_\gamma^2 \frac{R_{\rm g}}{r}  \left(1-\frac{R_{\rm g}}{r}\right)^{-1} \right]^{-1/2}
\end{equation}

\item We now rotate the direction of the photon over the polar angle of the magnetic field, to convert its direction (both polar and azimuthal) to a direction in a coordinate system with the magnetic field and the velocity along the polar axis. This then replaces $\mu_{\gamma{\rm s}}$ as defined above by a new angular variable $\mu_{\gamma v{\rm s}}$. We then perform a special relativistic transformation on the polar angle in this coordinate system to obtain the final angle between the photon and the magnetic field in the reference frame comoving with the outflow. This transformation is done through:
\begin{equation}
   \mu_{\gamma v 0} \; =\; \frac{\mu_{\gamma v{\rm s}} - \beta}{1-\beta \mu_{\gamma v{\rm s}}},
\end{equation}
where $\mu_{\gamma v 0}$ is the cosine of the angle between the photon and the velocity (or the magnetic field) in the local comoving frame, $\mu_{\gamma v{\rm s}}$ is the cosine of the angle between the photon and the velocity in the local stationary frame and $\beta=v/c$. Note that $v$ also has to be transformed to the local stationary reference frame first, as it is slightly modified by general relativity. The azimuthal angle is not modified by the special relativistic transformation.

\item The frequency of the photon can be converted straight from the global observer frame to the local comoving frame by computing the time component of its four-momentum in the local comoving frame from:
\begin{equation}
-P_{\rm co}^0 = -\frac{h \nu_{\rm co}}{c} = P^\sigma u_\sigma,
\end{equation}
where $h$ is the Planck constant, $\nu$ the photon frequency and the subscript co indicates a quantity in the local comoving frame. $P^\sigma$ is the covariant photon four-momentum and $u_\sigma$ is the contravariant four-velocity of the outflow. These four-vectors are given by:
\begin{equation}
   P^\sigma \; =\; \frac{h\nu}{c}\left(1\, ,\, \mu_\gamma \, ,\, \sqrt{1-\mu_\gamma^2} \, ,\, 0 \right)
\end{equation}
and
\begin{equation}
   u_\sigma \; =\; \gamma_{\rm Sch} \left(-\left(1-\frac{R_{\rm g}}{r}\right) ,\, \beta \mu_v \left(1-\frac{R_{\rm g}}{r}\right)^{-1} ,\, \beta \sqrt{1-\mu_v^2} ,\, 0 \right),
\end{equation}
where $\gamma_{\rm Sch}$ is the relativistic gamma factor in a Schwarzschild metric is determined using $u_{\sigma}u^{\sigma} =1$ to be
\begin{equation}
   \gamma_{\rm Sch} \; =\; \left[ 1 - \frac{R_{\rm g}}{r} - \beta^2 \mu_v^2 \left(1-\frac{R_{\rm g}}{r}\right)^{-1} - \beta^2(1-\mu_v^2) \right]^{-1/2}.
\end{equation}
In these equations $R_{\rm g} = 2GM/c^2$ is the gravitational radius of the neutron star and $\mu_v$ is the polar angle of the velocity in the global observer frame at infinity.

\item Having determined the direction and frequency of the photon in the comoving reference frame, we can pick a post-scattering direction in that same frame using Equation \ref{eq:kappa}. We will detail how we do this in Section \ref{sec:angleselect}. We do not change the frequency of the photon in the comoving frame since the scattering is in the Thomson regime, but its frequency will change in the global frame due to its new direction.

\item We transform the direction and frequency of the photon back to the global observer frame using the inverted functions of the transformations listed above.

\item Finally, the photon is assigned a new random travel distance in optical depth units, over which it can be moved in its new direction. This process is repeated until the photon is either destroyed by leaving the grid towards the star or the fireball, or until it escapes from the grid at the top.
\end{enumerate}

\subsection{Selecting scattering angles}
\label{sec:angleselect}

In principle the way to select a new polarization mode for a photon after scattering would be to pick a new mode using the probability for mode conversion, and then select a new angle from the probability density function for the post-scattering angle. This probability density function is given by:
\begin{align}
\label{eq:Ptheta}
P_i(\theta') =  \frac{{\rm d }\kappa_{ji}}{{\rm d}\theta'} = \frac{n_{\rm e}\sigt}{\rho} \sum_{\alpha=-1}^1 & \frac{\omega^2}{(\omega+\alpha\omega_{\rm C})^2 + \Gamma_{\rm e}^2/4} \left|e_\alpha^j \right|^2 \nonumber \\ 
 & \times \frac{3}{4} \left|e_\alpha^i (\theta') \right|^2 {\rm sin}\,\theta'.
\end{align}
Note that this equation is not normalized. The mode conversion probability can be computed from this equation, but only numerically. To pick a random angle $\theta'$ from this probability density function, we would have to compute the inverse function of its indefinite integral. This is again something that can only be done numerically. Instead of going through this process, we choose to pick post-scattering polarization modes and angles by trial and error. 

To pick a post-scattering mode and angle by trial and error, we generate a random mode $i_{\rm r}$ (1 or 2) and a random angle $\theta'_{\rm r} \in [0,\pi]$. We then compute $P_{i_{\rm r}}(\theta'_{\rm r})$ from Equation \ref{eq:Ptheta}, and compare this to a randomly generated test value in the range $[0,0.75\kappa_j]$. The value $0.75\kappa_j$ is an upper limit for the maximum value of $P_{i_{\rm r}}(\theta')$, which we use instead of the true maximum because that can only be computed numerically, while we already have a value for $\kappa_j$ from propagating the photon. We now compare our test value to $P_{i_{\rm r}}(\theta'_{\rm r})$. If $P_{i_{\rm r}}(\theta'_{\rm r})$ is greater than the test value we accept the values of $i_{\rm r}$ and $\theta'_{\rm r}$. If it is smaller we generate new random values and continue until we find a correct post-scattering mode and angle. This accept-reject method is the most computationally intensive part of our simulations, but is still quite manageable on a modern CPU.

\section{Results}
\label{sec:results}

Our simulations depend on a number of physical input variables as described in Section \ref{sec:model}. In Table \ref{tab:parspace} we give the baseline values we have used for these parameters in computing our results, and the range in which we have varied them to look for differences in the results.

\begin{table}
%\footnotesize
\caption{
Input variables of our simulations and their values.
}
\centering
\begin{tabular}{lll}
\hline\hline
Variable & Baseline value & Range \\
\hline
Diameter fireball 		& 20 km 			& 1-100 km \\
$B_0$ 				& $10^{15}$ G		& $10^{14}-10^{15}$ G \\
$v_\infty$ 				& 0.2 c 			& 0.01 - 0.99 c \\
$\rho_0$ 				& 0.01 g cm$^{-3}$ 	& $0.0001 - 0.1$ g cm$^{-3}$ \\
Blackbody temperature	& 10 keV			& 1-300 keV \\
\hline
\end{tabular}
\label{tab:parspace}
\end{table} 

\begin{figure}
\begin{center}
\includegraphics[width=\columnwidth]{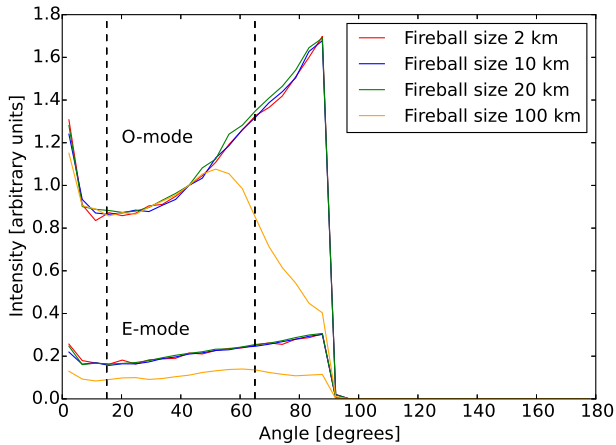}
\end{center}
\caption{Intensity of the escaping photon field as a function of angle with respect to the polar axis, for different fireball sizes. It is clear from this figure that the fireball size makes no significant difference to the beaming of the escaping photons for fireballs of size 2-20 km, but does change the intensity profile for the largest fireball.  Note that the angle (in this and all further figures) refers to the direction of the photon, not to the position from which it escapes. Our grid only runs from $0$ to $90$ degrees in polar angle, but in a physical system one would expect radiation to escape from the other side of the fireball as well (from the 'bottom' of the star), so that this intensity profile would be symmetrical around an angle of 90 degrees.  The dashed lines in this figure bracket the range of angles that would be sampled for the specific case of an observer inclination of $i=40^\circ$ and a magnetic pole offset of $\alpha = 25^\circ$.  See the text for more detail on how to relate the intensity shown in this figure (and the similar figures that follow) to the range of intensity in observed pulse profiles for other geometries.  
}
\label{fig:angle_fsize}
\end{figure}

We show our baseline result in Figure \ref{fig:angle_fsize}. This figure also shows results for different fireball sizes, which are not significantly different from the baseline result except in case of the largest fireball size, something we will discuss further below. This figure shows the intensity of the escaping radiation as a function of the direction in which it escapes with respect to the magnetic axis of the star, for the two different polarization modes. This intensity is calculated by dividing the number of escaped photons per angular bin by the size of that bin in units of solid angle. We call this the intensity profile of the escaping radiation. Note that this does not give any information about where on the star the radiation escapes, but merely about which direction it escapes in. It can be seen that this intensity profile is peaked in the direction parallel to the magnetic axis, as well as in the direction perpendicular to this axis. Note that this figure can be mirrored around 0 degrees, as that is the symmetry axis of our system. It can also be mirrored around 90 degrees, as at larger angles we would expect radiation coming from the other side of the star. As our model only considers one hemisphere, this radiation does not appear in our results.

This intensity profile can be related to an observed pulse profile by considering the polar angle between our line of sight and the magnetic axis (since the beam emerges at altitudes where light-bending effects can be neglected).  At an instant when this angle is 30$^\circ$, for example, we would receive the intensity shown for 30$^\circ$ in the intensity profile.  As the star rotates, this angle changes. The range within which it changes depends on the angle between the magnetic and rotation axes ($\alpha$), as well as on the inclination of the system ($i$). Only for a very `lucky' alignment will we see this angle move through the full 180$^\circ$ range of the intensity profile. In any other case, as the star rotates the angle between our line of sight and the magnetic axes changes by less than 180$^\circ$.  In general the angle will vary between $i-\alpha$ and  $i+\alpha$.  The observed pulse profile for an observer inclination of $i=40^\circ$ and a magnetic pole offset of $\alpha = 25^\circ$, for example, would sample the angle range $(15-65)^\circ$ shown in Figure \ref{fig:angle_fsize}, so that the intensity change as the star rotates would be relatively small.  The fluxes at the extremities of this range in Figure \ref{fig:angle_fsize} give an indication of the pulse fraction and the inter-pulse DC level flux. A pulse profile with more than one peak per rotation can arise if the angular intensity profile has multiple peaks within the sampled angle range, or if the angle between the magnetic axis and our line of sight crosses 90 degrees, as the emission from both hemispheres is the same in our model, so the intensity profile will be mirrored around 90 degrees.

In the rest of this paper, we will define the term intensity profile to mean the angle dependency of the escaping intensity in our simulations, as seen in Figure \ref{fig:angle_fsize} and similar figures further on. When we discuss the observed rotational modulations in the light curve, we will use the term pulse profile.

The fact that the fireball size does not significantly affect our results for the three smaller fireball sizes in Figure \ref{fig:angle_fsize} can be explained by the fact that all escaping photons are reprocessed well outside the fireball, deep inside the outflow zone, with opacity being controlled in particular by scattering at the electron cyclotron resonance, selected at the appropriate range of altitudes. The fact that we find no difference in the intensity profile of the escaping radiation also matches well with observations, which show that the pulse profile of the tail of the flare light curve is often stable over long periods of time \citep[see for example][]{Hurley1999a,Hurley2005a}, during which time the fireball is expected to shrink due to evaporation. We do see a significant difference for a fireball of size 100 km, which can be explained by the fact that here the fireball physically encroaches into the region where photons encounter the cyclotron resonance. It should be noted that 100 km is much larger than fireball sizes predicted in the literature, so this outlying result likely has little bearing on real magnetars.

The angular distribution of the intensity is principally governed by the radial velocity profile and the dipolar flaring of the magnetic field.  As the polarization dependence of the opacity couples with the angular directions of photons with respect to the local {\bf B} direction, the simulation naturally generates different angular intensity profiles for the O-mode and the E-mode.

Several of our other parameters merely have the effect of changing the fraction of photons that escape, and the distribution of those photons between the two modes, but not the intensity profile. This is the case for the magnetic field strength, the density and the blackbody temperature of the input spectrum. The one parameter that we do find to change the pulse profile of the escaping radiation significantly is the velocity. We will discuss this dependence separately in Section \ref{sec:vdep}. Effectively the field strength, density and temperature just change the total optical depth of the system. This changes what fraction of all photons escape, how likely they are to convert between modes, and the relative importance of photon diffusion to photon advection. It also changes the outcoming spectrum, which we will discuss in Section \ref{sec:specres}. However, the total optical depth of the system does not significantly change the intensity profile.

This means that observations of the pulse profile of a magnetar flare cannot be used to identify the density and magnetic field strength of the system in our model. However, an improved model that computes the properties of the outflow self-consistently from the radiation field, i.e., incorporating the influences of Compton drag on the wind dynamics, would provide constraints on these properties.  The angular dependence of the photon polarization degree, and how it changes over time as the properties of the system change, is potentially observable with future X-ray polarimetry missions \citep{Soffitta2013a, Weisskopf2014a, Dong2014a, Jahoda2015a}.  Our results also show that in general the O-mode is beamed slightly more strongly than the E-mode. This should manifest itself as distinctive phase-dependent signatures of both polarization degree and position angle in the tail of giant flares. More complete modeling of the radiation/wind interaction and opacity at higher altitudes is needed in future work to fully leverage the polarimetric data such instruments could provide.

\subsection{Velocity dependence}
\label{sec:vdep}

\begin{figure}
\begin{center}
\includegraphics[width=\columnwidth]{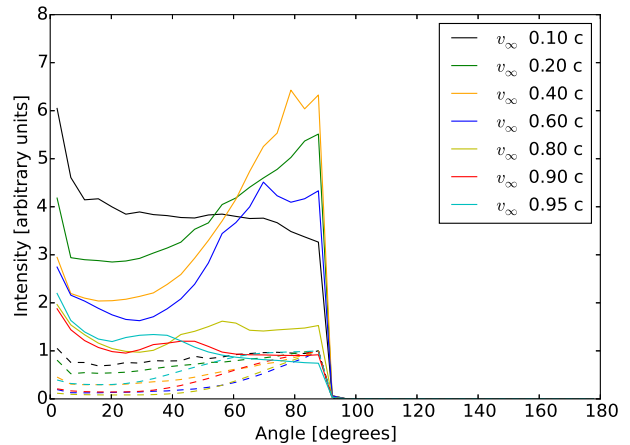}
\end{center}
\caption{Angular intensity profile of the escaping radiation for different outflow velocities, for both polarization modes. Solid lines indicate the O-mode profiles, while dashed lines indicate the E-mode. All results have the same mass loss rate, which has been accomplished by scaling the density at the base of the outflow. These profiles are normalized with respect to the E-mode intensity in the last angle bin before 90 degrees, which is why the E-mode curves intersect there. The O-mode intensities are normalized to the same E-mode intensity values, so the ratio between the E-mode and the O-mode intensity for one outflow velocity is preserved. It can be seen that this E-mode to O-mode ratio varies significantly. 
}
\label{fig:angle_v}
\end{figure}

\begin{figure}
\begin{center}
\includegraphics[width=\columnwidth]{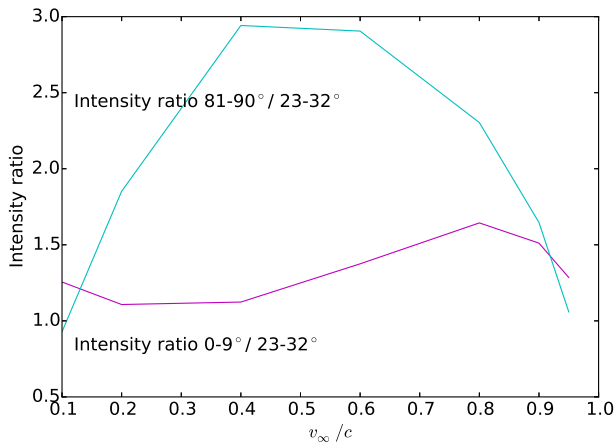}
\end{center}
\caption{ Plot of the ratio between the intensity at small angles and at moderate angles, as well as the ration between the intensity at large angles and at moderate angles. These intensities are for the E- and O-mode combined. The reference angle range of 23-32$^\circ$  corresponds to the lowest average intensity over all velocities, which is why we use this angle range for comparison. It can be seen that the intensity between small and moderate angles is fairly constant, while the ratio between large and moderate angles is a strong function of velocity. These results have been computed at the same mass loss rate of $\sim10^{18}$ g s$^{-1}$.
}
\label{fig:beaming_v}
\end{figure}

We find that the velocity of the outflow is the main factor that determines the angular intensity profile of the escaping radiation. Figure \ref{fig:angle_v} shows this profile for different values of $v_\infty$, the velocity of the outflow at infinity. This figure shows that the shape of the angular intensity profile is a strongly varying and non-monotonous function of v. We quantify this observation if Figure  \ref{fig:beaming_v}, by showing the ratio between small, moderate and large angle intensities.  Additionally, this figure shows a strong velocity dependence in the ration between the O-mode and E-mode intensities, with higher velocities giving a larger fraction of the total intensity in the E-mode.

Figure \ref{fig:beaming_v} shows that the intensity at small angles is always a bit higher than the minimum intensity, and that this slight beaming along the magnetic axis is not a strong function of velocity. Additionally, it shows that the intensity at large angles is not always larger than the minimum intensity, and that this ratio is a strong function of velocity, peaking at an intensity ratio of around three at intermediate velocities. This suggests that two different mechanisms are responsible for these two peaks, one of which does depend on velocity, while the other one most likely does not. Additionally, there appears to be a third peak present in Figure \ref{fig:angle_v}, which first occurs around 80 degrees for the result with $v=0.4c$, and then shifts to smaller angles for higher velocities. This peak does neither disappear nor move significantly for results created in less or more CPU time, suggesting there is indeed a third, velocity-dependent beaming mechanism at work.

The highest ratio between the peak and minimum intensity in our models is about a factor three. This is somewhat higher than (but fairly close to) the ratio between pulse and off-pulse emission observed in intermediate flares, which is typically around a factor two \citep[see for example][]{Ibrahim2001a, Woods2005a, Kaneko2010a}, while in giant flares this ratio can be as large as an order of magnitude \citep{Hurley2005a}. As noted in Section \ref{sec:results}, the observed pulse profile will nearly always be shallower than our simulated intensity profile, due to inclination effects. Thus, the relative amplitude of our intensity profiles of a factor three corresponds quite well to an observed pulse profile with a relative amplitude of a factor two.

The intensity profiles in Figure \ref{fig:angle_v} all have fairly wide peaks, which may match observations of intermediate flares, such as the intermediate flare observed by \citet{Ibrahim2001a}, while observations of giant flares often show much narrower peaks such as the giant flares observed by \citet{Feroci2001a} and \citet{Hurley2005a}. Combined with the higher amplitudes seen in the giant flares, this suggests that our simple model is not able to reproduce the strong beaming of the giant flares, while it can approximate the results of the intermediate flares. 

The applicability of our model to intermediate flares is somewhat debatable, as these flares may not be energetic enough to create a fireball and ablate a significant amount of matter from the star, and their luminosities may not always be high enough create an outflow. However, our results show some degree of beaming along the magnetic axis regardless of the velocity of the outflow or the size of the fireball, which suggests that this beaming does not depend on the presence of a fireball and an outflow, but may also occur for another source of radiation as long as there is matter present in the atmosphere to scatter the radiation.

\subsection{Spectral variations.}
\label{sec:specres}

\begin{figure}
\begin{center}
\includegraphics[width=\columnwidth]{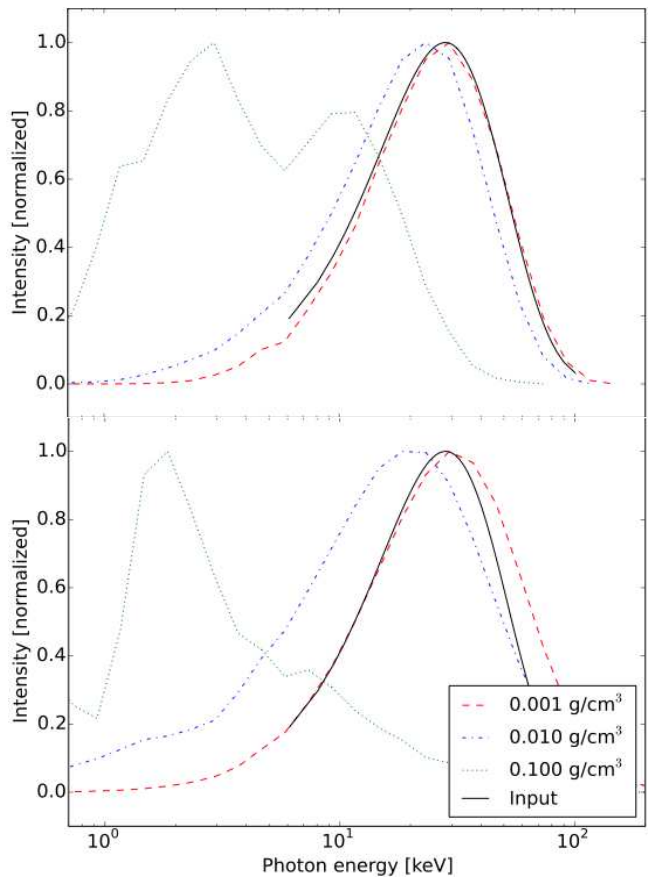}
\end{center}
\caption{ Normalized spectra for three different values of $\rho_0$, the density at the base of the outflow, compared to the input spectrum, shown for two different values of the outflow velocity: $v_\infty=0.2$ c (top panel) and $v_\infty=0.8$ c (bottom panel). This shows that the average energy of the output spectrum becomes lower for higher densities (because of the larger optical depth), and higher for higher velocities. As these effects work in opposite direction, the average energy of the output spectrum can be both higher or lower than that of the input spectrum. The three densities shown give an optical depth to Thomson scattering through the system of $140$ for a base density of $10^{-3}$ g cm$^{-3}$, and factors of 10 and 100 higher for the higher densities. This gives an estimate of the optical depth through the system for O-mode photons, while E-mode photons have lower optical depth. The irregular shapes of the spectra for the highest density are signatures of poor statistics --- see text.
}
\label{fig:spectrum_density}
\end{figure}

In addition to looking at the intensity profiles of our escaping emission, we also look at the spectrum. We do not attempt to create an accurate model of a magnetar spectrum, but we do look at shifts of the entire spectrum to higher and lower energies. We find that the average energy of our output spectrum is strongly influenced by the total optical depth through our model, as influenced by the density, magnetic field strength and photon temperature. Figure \ref{fig:spectrum_density} shows the output spectrum for three different values of the density at the base of the outflow $\rho_0$. We find that higher densities lead to softer spectra.

There are three main effects that shift the average energy of the output spectrum with respect to the input spectrum, working in different directions. The main effect shifting the spectrum towards lower energies is the fact that lower energy photons have lower optical depth in the E-mode, as the E-mode opacity scales roughly as $\omega^2/\omegaB^2$. This causes lower energy photons to escape mores easily, and this effect is more pronounced when the optical depth of the entire system is larger, such as at high densities. This does not just affect the E-mode intensity, as photons regularly convert between the two polarization modes. The main effect shifting photons to higher energies is the presence of (mildly) relativistic electrons, which on average cause photons to gain energy.   Which of these two effects is stronger depends on the specific parameters of the model, so that some of our output spectra have higher average energy than the input spectrum, while others have lower average energy. This is illustrated in Figure \ref{fig:spectrum_density}.

The final main energy shift in our results is caused by the fact that for different velocities and different optical depths, the distribution of escaping photons over the two polarization modes and over angles is different. This causes the overall spectrum to shift in a different way from the spectrum for a single mode in a single direction. This is what causes the high density spectrum to shift to lower energy when going from the top to the bottom panel in Figure \ref{fig:spectrum_density}, while the low density spectrum shifts to higher average energy.

Observe also the irregular shapes of the spectra for the highest density.  These are caused by poor statistics, due to the fact that for these high optical depths very few photons escape, as the chance of escape for a single photon on a random path falls off exponentially with increasing optical depth. Thus, these irregular shapes should not be taken to have any physical meaning, and we will only discuss the average energy of these spectra as a whole.
 
In addition to these main effects, a small energy shift may be introduced by the angle anisotropy of the scattering cross section, as the scattering angle determines how the energy of the photon changes in transforming it from the global reference frame to the local comoving frame and back. A similar effect may be introduced by the curvature of the magnetic field, which causes the angle a photon makes with respect to the outflow direction not to be preserved as the photon propagates outwards.  

\begin{figure}
\begin{center}
\includegraphics[width=\columnwidth]{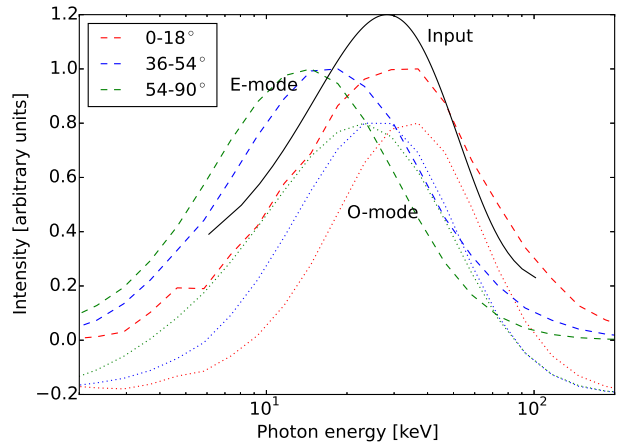}
\end{center}
\caption{ Normalized spectra for three different angle bins, for $v_\infty=0.8c$. The vertical offset is artificial to aid readability. Dashed lines correspond to the E-mode spectrum, while dotted lines correspond to the O-mode. This figure shows that photons escaping in directions close to parallel to the polar axis have a somewhat higher mean energy than those escaping close to perpendicular, and that O-mode photons are slightly more energetic than E-mode photons.
}
\label{fig:spectrum_angle}
\end{figure}

In addition to looking at the output spectrum as a whole, we can also create output spectra for specific angular directions. Figure \ref{fig:spectrum_angle} shows the output spectrum for three different angle bins for the same simulation. This figure shows that photons escaping at narrow polar angles have higher energy than photons escaping at wide polar angles. In observational terms, this means that we predict the mean photon energy of the on-pulse emission to be different from the mean energy of the off-pulse emission, with the on-pulse emission to have higher energy if the observed beaming peak lies along the magnetic axis.

Interestingly, more energetic on-pulse than off-pulse emission is exactly what was observed by \citet{Boggs2007a} for the 2004 giant flare of SGR 1806-20, while \citet{Feroci2001a} found exactly the opposite for the 1998 giant flare from SGR 1900+14. However, it should be noted that \citeauthor{Feroci2001a} measure a (100-700)/(25-150) keV hardness ratio from {\it BeppoSAX} and {\it Ulysses} data, while \citeauthor{Boggs2007a} measure a best-fit blackbody temperature from 20-100 keV {\it RHESSI} data, finding temperatures from 5 to 12 keV. These differing energy ranges suggest that the results from \citeauthor{Boggs2007a} are more likely to be applicable to our model. The \citeauthor{Feroci2001a} results may be caused by a process at higher energies that is not included in our model. Alternately, if most of the photons are around or below 25 keV, a higher mean photon energy may actually cause a lower hardness ratio, as more photons shift into the lower band while the higher band remains relatively unchanged.

In principle it could also be possible to test our predictions for spectral variations with rotational phase for short bursts, as phase resolved spectroscopy data exist for these bursts \citep{Younes2014a, Collazzi2015a}. However, these datasets incorporate many different bursts, so one would have to make some assumptions about the properties of the fireball and the outflow between different bursts. Additionally, being of lower fluence than large and giant flares, these bursts might not create a fireball or an outflow, which would make some changes to our model necessary.

\section{More complicated geometries}
\label{sec:complicated}

The model we have considered so far has a very simple geometrical setup, with a purely dipolar magnetic field and an outflow with the same properties everywhere outside the fireball.  So far, we have shown that this model cannot explain the observed pulse profiles of giant flares, as those have higher amplitudes and narrower peaks than our simulated intensity profiles. However, the assumptions of our simple model may have oversimplified the problem. In this section, we test whether a slightly more complicated model can create the pulse profiles observed in giant flares. 

Firstly, a magnetar may well have a significant multipolar field component, which would be the only way to trap a fireball with a shape different from a torus, and also the only way to get more than two peaks in the pulse profile. The model proposed by \citet{Thompson2001a} for a four peaked pulse profile involves multiple fireballs close together, with a narrow fan beam of radiation escaping from in between these fireballs. Secondly,  our assumption that the outflow has the same properties everywhere outside the fireball is flawed, as most of the magnetic field lines outside the fireball are closed, and only a small fraction of field lines open out to infinity. On closed field lines matter will pile up, making a relativistic outflow along those field lines impossible.

In this section we make two adaptations to our model to test the effects of changing these two assumptions of our model. In Section \ref{sec:narrowflow} we adapt our model to simulate a relativistic outflow with a small opening angle, with matter in between this outflow and the fireball piling up on closed field lines. In Section \ref{sec:narrowregion} we introduce a solid boundary with a small opening angle into our models to simulate the effect of radiation escaping from two fireballs that are close together. 

\subsection{A narrow relativistic outflow above the polar cap}
\label{sec:narrowflow}

In a realistic magnetar model, most of the matter ablated from the surface will end up on closed field lines, as only a small fraction of field lines open out to infinity. Matter on closed field lines will be able to flow out initially, but should pile up when it meets matter coming from the other side of the star (assuming a symmetrical flare), or when it physically hits the other side of the star. Thus, the relativistic outflow we have assumed in our model so far is only feasible inside a limited bundle of field lines, which starts as a cone with a narrow opening angle directly above the polar cap and gradually diverges to a wide bundle of field lines opening out to infinity. The distance at which field lines start opening out to infinity can be estimated from the trapping luminosity given by \citet{Lamb1982}:
\begin{equation}
L_{\rm tr} \simeq 2.1\times10^{49} \left(\frac{B}{B_{\rm cr}}\right)^2\left(\frac{r}{10\,{\rm km}}\right)^2 {\rm erg}\,{\rm s}^{-1}.
\end{equation}
This is the luminosity needed for radiation force to open out the field lines, which is the relevant force when considering a radiation driven wind. Scaling from this to typical magnetar flare parameters we find that the field lines beyond $10^3-10^4$ km will be opened out by a radiation driven outflow. From this, we use the classical equation for the shape of a dipolar field line ($r = r_{\rm max}\,{\rm sin}^2\theta$) to estimate the opening angle at the surface of the bundle of field lines that open out to infinity. This gives an opening angle at the surface in the range of 2-6 degrees.

\begin{figure}
\begin{center}
\includegraphics[width=\columnwidth]{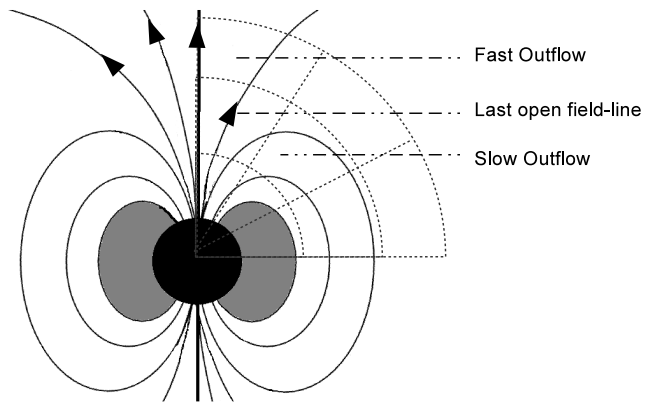}
\end{center}
\caption{ Geometry of the adaptation to our model made in Section \ref{sec:narrowflow}. Instead of having a (mildly) relativistic outflow everywhere outside the fireball, we model a fast outflow on narrow bundle of field lines, assumed to be only those field lines that open out to infinity, while having a slow outflow everywhere else, to simulate matter piling up on closed field lines. This figure is a copy of Figure \ref{fig:geometry}, and the labels in that figure also apply to this figure. 
}
\label{fig:geometry_advanced}
\end{figure}

We now adapt our model to have a relativistic outflow only inside a bundle of field lines that opens out to infinity.  This adaptation is illustrated in Figure \ref{fig:geometry_advanced}.  We do this by defining a radius $r_{\rm open}=10^8$ cm beyond which we assume all field lines open out. This radius defines a bundle of field lines which starts as a narrow cone close to the surface, and opens out to cover a complete hemisphere beyond $r_{\rm open}$. Inside this bundle, we create a relativistic outflow with a final velocity $v_\infty=0.8$ c. Outside this bundle we reduce this velocity by a factor $dv=80$ to 0.01 c, simulating matter  slowly moving out as it piles up on a closed field line loop. We increase the density outside the open bundle by the same factor, to keep the mass loss rate constant everywhere outside the fireball, as the amount of matter ablated should not depend on whether the field line it ends up on is open or closed. We also compute solutions with $r_{\rm open}=10^9$ cm, and with a velocity reduction factor $dv$ of 10 instead of 80, to explore the effects of our assumptions. This model setup is closer to the beaming scenario set out by \citet{Thompson2001a} than our basic model, as they assume a narrow jet of X-ray radiation and ablated matter, rather than a widely spread outflow, with ablated matter outside the jet forming a sort of nozzle.

\begin{figure}
\begin{center}
\includegraphics[width=\columnwidth]{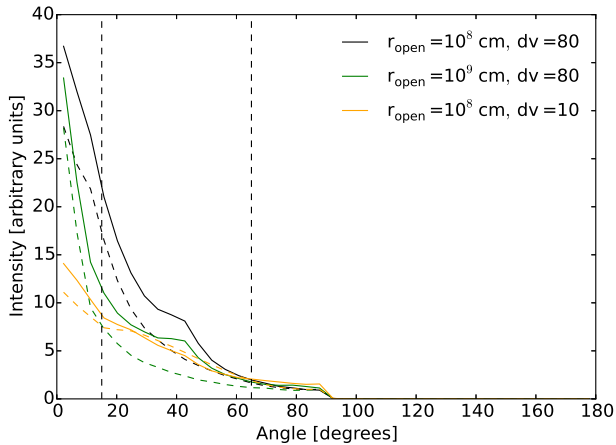}
\end{center}
\caption{Intensity profile of a model where there is only a relativistic outflow inside the bundle of field lines that opens out to infinity. The radius at which we assume field lines open out to infinity is either $10^8$ or $10^9$ cm (see figure legend). Outside this bundle of field lines, the velocity is lower by a factor $dv$, which is either 80 or 10 (see legend), and the density is higher by the same factor. Solid lines indicate the O-mode, while dashed lines indicate the E-mode. This simulates a more realistic outflow situation, where only matter that is on field lines that open out to infinity can form a relativistic outflow, while matter on closed field lines piles up.  As in Fig.~2, the dashed vertical lines here bracket the range of angles that would be sampled for the specific case of an observer inclination of $i=40^\circ$ and a magnetic pole inclination of $\alpha = 25^\circ$ to the spin axis.  This setup gives much stronger beaming than our basic model, as the ratio of maximum to minimum intensity is above 30 for the two models with $dv=80$, and above 10 for the model with $dv=10$.
}
\label{fig:angle_funnel}
\end{figure}

 Figure \ref{fig:angle_funnel} shows the results of this adaptation of our model for three different combinations of $r_{\rm open}$ and $dv$. It is clear that this model is capable of creating much stronger beaming than our basic model, as well as a more narrow beam. The ratio between maximum and minimum intensity is well above the order of magnitude or so seen in giant flares. This means our model is able to recreate the peak/off-peak flux ratio observed in giant flares, even for less than ideal inclination angles. Additionally, for large $r_{\rm open}$ the intensity profile we find is very sharply peaked, enabling the sharply peaked pulse profiles observed in giant flares.

\subsection{A narrow scattering region between two fireballs}
\label{sec:narrowregion}

Magnetars may well have a magnetic field with significant multipolar components, rather than a simple dipole structure. \citet{Thompson2001a} sketch a picture where two sphere-like fireballs trapped close together in a multipolar field create a fan beam of radiation. In this section we adapt our basic model to simulate a situation where two fireballs are close together. We do this by simply creating a hard boundary in our model at an opening angle of 10 degrees, from the surface of the star up to a height of 20 km above the surface, which is equal to the diameter of the fireball we use as our base value throughout our models. While this does not simulate a multipolar field, it does simulate the effect of having two physical objects close together on the surface of the star, with only a narrow area in between for radiation to escape from.

We find that the results from this adaptation of our model are identical to those of our base model. Just as in the case of different fireball sizes, changing the physical structure near the star has no influence on the intensity profile coming out, as nearly all escaping photons scatter multiple times well above the fireball. Thus, creating a narrow nozzle close to the star has no influence on the intensity profile if this nozzle does not extend to large distance from the star, which would require a fireball of much greater volume than any estimates suggest.

This does not mean that a multipolar field structure would have no effect on the escaping intensity profile, as a true multipolar field setup would also create a very different outflow from a dipolar field. Changing the geometry of the magnetic field and the outflow would be the only way to create a pulse profile with more than one strongly beamed peak per hemisphere in our models. This would require either multiple distinct open field line bundles, or some sort of fan beam bundle of open field lines.

\section{Discussion and Conclusions}
\label{sec:conclusion}

Our simulations are able to reproduce the basic properties of the rotational modulation seen in the tails of magnetar flare light curves. Summarizing, our basic model consists of a torus-shaped fireball trapped in a dipolar magnetic field with a (mildly) relativistic outflow outside this fireball. We scatter photons through this system, starting from the base of the fireball and letting them escape when they get sufficiently far away from the star. We take Compton scattering off free electrons to be the only form of opacity, but do include special and general relativistic light bending effects in our model.

The angular intensity profile of our basic model only depends strongly on the velocity of the outflow. The size of the fireball has no influence on this profile as long as the fireball is not so unrealistically large that it intrudes into the cyclotron resonance region. This matches well with observations, as most modulations in magnetar flare light curves show very little change in the pulse profile over time even though the fireball should be shrinking.  Additionally, we find that the density, magnetic field strength and seed photon temperature only influence the distribution of photons between the two polarization modes, as these parameters all just have the effect of changing the total optical depth of the system, and for the magnetic field strength and seed photon temperature the ratio between the optical depth to E- and O-mode photons. A higher optical depth of the entire system causes a larger intensity in the E-mode compared to the O-mode.

The velocity of the outflow does strongly influence the shape of the angular intensity profile, which shows a small degree of enhancement along the magnetic axis at all velocities, as well as beaming perpendicular to the magnetic axis that is strongest for velocities around $0.5 c$ and disappears at low and high velocities. This suggests that the beaming along the magnetic axis is a property of the general setup of our outflow and magnetic field structure, and the Compton scattering opacity in a very strong magnetic field, while the beaming perpendicular to the magnetic axis is directly caused by the outflow.

The highest ratio between the maximum and minimum intensity in our basic model is around three. This ratio can in principle be used to explain any observed pulse profile with a ratio between maximum and minimum intensity of three or less, as the observed ratio depends strongly on the angle between the rotation and polar axes of the star as well as the inclination at which we observe it. Thus, our simple beaming model can adequately explain the pulse amplitude of most intermediate flares that show rotational modulation in their light curve. We do not find very narrow peaks of radiation in our basic model, but as the pulse profiles of the modulation in intermediate flares are usually fairly wide \citep[see for example][]{Ibrahim2001a,Woods2005a} our intensity profiles should be able to just about explain the shapes of these pulse profiles.

Whether our models are applicable to short and intermediate flares is debatable, as these flares do not necessarily have the energy to create a fireball and an outflow.  However, the results of our simple model show no dependence on the size of the fireball, and show that some degree of beaming along the magnetic axis can be created independently of the properties of the outflow. We thus conclude that the geometry of the magnetic field generates beaming along the magnetic axis as long as there is sufficient matter present outside the fireball to scatter the radiation. However, this scattering matter most likely has to come from some form of outflow. This outflow would not necessarily have to be continuous, as long as the ejected matter does not have time to fall back to the surface. 

In the case of giant flares, our basic model clearly does not suffice to explain the observations, both in terms of the amplitude and in terms of the shape of the pulse profile. However, we find that with a simple adaptation these problems are alleviated. We adapt our model to have a relativistic outflow along only those magnetic field lines that open out to infinity, rather than everywhere outside the fireball. This is also closer to physical reality, as a relativistic outflow along closed field lines cannot be sustained, as matter would quickly pile up. Using this model we find a much larger ratio between the maximum and minimum intensity, and a narrower angular intensity profile, both of which vary strongly with the properties assumed for the relativistic outflow. This makes it possible to recreate the pulse/off-pulse ratio of the giant flares of up to an order of magnitude without needing a particularly favourable inclination axis. The angular intensity profile of this model can also be sufficiently narrow to create the sort of sharp peaks seen in the giant flare pulse profiles \citep{Feroci2001a,Hurley2005a}.

 Unlike our simple model, which shows some beaming regardless of the outflow velocity, our adapted model does require a significant outflow, as that is the only natural way to create a distinction between a narrow open field-line bundle and the rest of the atmosphere. This leads us to the conclusion that giant flares require a radiation driven outflow during the entire tail of the light curve to explain the observed modulations.

Our simulated intensity profiles can only be made into a prediction of the observed pulse profile by assuming values for the inclination angle and the angle between the magnetic axis and the rotation axis of the system. It has been shown that these angles can be strongly constrained by modelling the pulse profile of the quiescent emission of a magnetar \citep[see][]{Perna2008a,Bernardini2011a}. Combining this analysis of the quiescent emission with our model for the flare emission for the same source would greatly improve the descriptive power of our model, and would enable us to fit the properties of the outflow in our model to the observed light curve. 

Having shown that the angular intensity profile of the escaping radiation depends mostly on the chosen size and properties of the relativistic outflow,  we can now discuss how our model could be adapted to enable the more complicated four-peaked pulse profile seen in the 1998 giant flare \citep{Feroci2001a}. Our results show that structure close to the surface of the star does not enable a more complicated pulse profile, as the radiation is always reprocessed higher up in the atmosphere, where the optical depth to E-mode photons is highest (because of the scaling of the opacity with magnetic field strength).  Thus, we conclude that the way to make a more complicated pulse profile would be by creating a geometrically different outflow, such as a fan beam, multiple outflows per hemisphere, or simply an outflow that is not spherically symmetric. This is not possible in a dipole magnetic field, but should be possible by introducing multipolar components, as a more complex field structure would change the shape of the field-line bundles that open outwards. It could also create multiple fireballs, causing matter to ablate from multiple sections of the surface and form an outflow with an asymmetric density profile. 

While making an accurate model of the spectrum of a magnetar flare is outside the scope of this paper, we can draw some conclusions from the average energy of the spectra our model produces. Firstly, we find that a higher optical depth of our system leads to a softer spectrum. This observation may be of value in interpreting spectral changes during the tail of a magnetar flare. Additionally, we find that the emission escaping parallel to the polar axis  is harder than the emission escaping perpendicular to the polar axis. This means that for the beaming scenario of our more realistic adapted model, which shows strong beaming along the polar axis, the hardness of the spectrum observed from magnetar flares should correlate with the pulse profile. This prediction matches the observations of the 2004 giant flare from \citet{Boggs2007a}, who show that the best fit blackbody temperature of the phase-folded emission correlates with the phase-folded light curve. Surprisingly, \citet{Feroci2001a} find that for the 1998 giant flare the (100-700)/(25-100) keV hardness ratio is anti-correlated with the pulse profile. We believe this apparent contradiction can be explained by the fact that the lower boundary of this hardness ratio of 25 keV falls fairly close to the peak of a typical magnetar spectrum. Therefore, the spectrum shifting to higher energy can actually cause a lower hardness ratio in this case, by shifting more counts into the denominator of the ratio without significantly affecting the numerator.

Our results show only a moderate difference between the angular intensity profile between the two polarization modes. However, the distribution of photons between the two polarization modes is strongly variable with the various physical parameters (velocity, density, seed photon temperature, magnetic field strength) of our model. This means that future X-ray polarimetry missions, such as {\it XIPE} \citep{Soffitta2013a}, {\it IXPE} \citep{Weisskopf2014a}, {\it XTP} \citep{Dong2014a} and {\it PRAXyS} \citep{Jahoda2015a} could be able to add significant value to magnetar flare observations, by tracking what happens to the polarization signature during the tail of a magnetar flare. Because the photon mode distribution may change while the pulse profile does not, this would add valuable constraints to any attempt to fit magnetar flare light curves using the model set out in this paper.

In conclusion, we find that the model for magnetar flare beaming set out in this paper, as based on the model proposed by \citet{Thompson2001a}, works well to simulate the rotational modulations observed in these flares. We believe this model is a promising way to fit light curves of the tails of magnetar flares, as such fits will provide constraints on the field structure of the magnetar and the properties of the expected relativistic outflow. These constraints can be made stronger by including  the results of fits to the quiescent emission, as well as  changes in the spectrum and the polarization of the observed radiation. 

Note: Whilst our paper was in review, an offering addressing this topic by \citet{Yang2015}, now in press, was posted to the arXiv.  Their paper focuses on potential polarization signatures of the trapped fireball that forms in the aftermath of the giant flare.   The set-up of their model is however quite different to the canonical fireball beaming model of \citet{Thompson1995, Thompson2001a}  explored in our study. In particular it does not include the baryonic component ablated from the stellar surface by the super-Eddington luminosities. In the canonical model it is this component that controls the evolution and spectrum of the fireball, and which gives rise to the scattering material and outflowing jet necessary to collimate the outgoing radiation into a sharp beam.  As such it is not clear whether the \citet{Yang2015} model can generate the observed strong pulse profiles. 

\section*{Acknowledgements}

TvP and RAMJW acknowledge support from the ERC through Advanced Grant no. 247295. ALW acknowledges support from NWO Vidi Grant no. 639.042.916. We thank SURFsara (www.surfsara.nl) for the support in using the Lisa Compute Cluster.  MGB acknowledges support from NASA {\it Fermi} Guest Investigator grant NNX13AP08G and National Science Foundation grant AST-1517550.  We would also like to thank Roberto Turolla for helpful comments.

%%%%%%%%%%%%%%%%%%%%%%%%%%%%%%%%%%%%%%%%%%%%%%%%%%

%%%%%%%%%%%%%%%%% APPENDICES %%%%%%%%%%%%%%%%%%%%%

\appendix

\section{Cyclotron Decay Widths}

As indicated in Section~\ref{sec:scattering}, to provide accuracy in the opacity at 
frequencies in and near the cyclotron resonance $\omegaB$, full QED character of 
the cyclotron decay widths in the Lorentz profile must be treated.
The spin-averaged width/rate $\Gamma_{\rm e}$ for cyclotron decay 
of the virtual electron from the $n=1$ first excited Landau state
is employed in the fully relativistic magnetic Compton physics formalism 
of Gonthier et al. (2014).  It  assumes the compact form
\begin{equation}
   \Gamma_{\rm e} \;\equiv\; \frac{ \fsc\, \omegaB}{ {\cal E}_1} 
   \int_0^{\Phi} \frac{d\kappa \, e^{-\kappa}}{
        \sqrt{(\Phi -\kappa )\, (1/\Phi - \kappa )}}\;
        \Biggl\lbrack 1 - \frac{\kappa}{2} \biggl( \Phi +
                \frac{1}{\Phi}\, \biggr)\, \Biggr\rbrack \;\; ,
 \label{eq:Gamma_ave}
\end{equation}
where the integration is over the angles of the cyclotron photons. 
Also $\Phi = (\sqrt{1+2 B/B_{\rm cr}}-1)/(\sqrt{1+2B/B_{\rm cr}}+1)$.  This
expression is from Eq.~(14) of Baring, Gonthier \& Harding (2005);
the cyclotron width for an electron with momentum $p_z$ parallel to {\bf B}
is given in Eq.~(13) therein.  The factor ${\cal E}_1= \sqrt{1+(p_z/m_ec)^2+2B/B_{\rm cr}} \to 1+B/B_{\rm cr}$ is evaluated precisely 
at the cyclotron energy $\omega =\omegaB$.  In the Compton scattering problem, 
$p_z$ for the intermediate state equals the incoming photon frequency (in dimensionless 
units), i.e., $p_zc\to \hbar \omegaB$, and the virtual electron receives a kick along {\bf B} from the initial photon.  Accordingly, 
one needs to include the ${\cal E}_1$ factor as a time dilation correction.  It is 
small for sub-critical fields, but is a large influence on the width of the resonance for 
super-critical fields --- the decay time is lengthened, and so the width is reduced.

For deployment in a CPU-intensive simulation, it is computationally expedient to approximate 
the width integral empirically.  The two asymptotic limits
for this spin-averaged $n=1\to 0$ cyclotron 
width are $\Gamma_{\rm e}\approx 2e^2\omegaB^2 /(3m_{\rm e}c^3)$ when $B\ll B_{\rm cr}$,
and $\Gamma_{\rm e}\approx \fsc\, [m_ec^2/\hbar]\, (1-1/e)$ when $B\gg B_{\rm cr}$; see Baring, Gonthier \& Harding (2005) 
and Gonthier et al. (2014) for expanded discussions of these.  Leveraging these asymptotic forms, we define
\begin{equation}
   \gamma_l \; =\; \frac{2}{3} \left( \frac{B}{B_{\rm cr}} \right)^2
   \quad ,\quad
   \gamma_h \; =\; 1 - \frac{1}{e}\;\approx\; 0.632\quad ,
 \label{eq:asymp_forms}
\end{equation}
so that $\Gamma_{\rm e}\approx \fsc \,[ m_ec^2/\hbar ]\, \gamma_{l,h}$ for very low and sufficiently high fields.  
A useful empirical approximation is then 
\begin{equation}
   \Gamma_{\rm e} \; \approx\; \fsc\, \frac{m_ec^2}{\hbar}\,
      \frac{\gamma_l\, \gamma_h}{\Bigl[ (\gamma_l)^{1/q} + (\gamma_h)^{1/q} \Bigr]^q}
   \quad ,\quad
   q \; =\; \frac{2\pi}{3}\quad .
 \label{eq:width_empirical}
\end{equation}
This choice of $q$ yields a precision of better than around 2\% when compared 
with the exact form for $\Gamma_{\rm e}$ in Eq.~(\ref{eq:Gamma_ave}).  
The accuracy is worst at $B\sim 4B_{\rm cr}$, and is excellent for $B\ll B_{\rm cr}$, and $B\gg B_{\rm cr}$.
The empirical formula in Eq.~(\ref{eq:width_empirical}) is the width that is adopted for 
the opacity expression in Eq.~(\ref{eq:kappa}), used to generate all simulation results
for the paper.

% Don't change these lines
\bsp	% typesetting comment
\label{lastpage}
\end{document}